\documentclass[apj,graphicx]{emulateapj}
\usepackage{apjfonts}

\usepackage{graphicx}

\newcommand\cxo{{\it Chandra}}
\newcommand\mxb{MXB~1659$-$29}
\newcommand\ks{KS~1731$-$260}

\shortauthors{Cackett et al.}
\shorttitle{MXB~1659$-$29 in quiescence}

\begin{document}
\title{A change in the quiescent X-ray spectrum of the neutron star low-mass X-ray binary MXB~1659$-$29}

\author{E.~M.~Cackett\altaffilmark{1}}
\author{E.~F.~Brown\altaffilmark{2}}
\author{A.~Cumming\altaffilmark{3}}
\author{N.~Degenaar\altaffilmark{4}}
\author{J.~K.~Fridriksson\altaffilmark{5}}
\author{J.~Homan\altaffilmark{6}}
\author{J.~M.~Miller\altaffilmark{4}}
\author{R.~Wijnands\altaffilmark{5}}

\email{ecackett@wayne.edu}

\affil{\altaffilmark{1}Department of Physics \& Astronomy, Wayne State University, 666 W. Hancock St, Detroit, MI 48201, USA}
\affil{\altaffilmark{2}Department of Physics \& Astronomy, National Superconducting Cyclotron Laboratory, and the Joint Institute for Nuclear Astrophysics, Michigan State University, East Lansing, MI 48824, USA}
\affil{\altaffilmark{3}Department of Physics, McGill University, 3600 rue
University, Montreal QC, H3A 2T8, Canada}
\affil{\altaffilmark{4}Department of Astronomy, University of Michigan, 500
Church St, Ann Arbor, MI 48109-1042, USA}
\affil{\altaffilmark{5}Astronomical Institute `Anton Pannekoek', University of
Amsterdam, Science Park 904, 1098 XH Amsterdam, the Netherlands}
\affil{\altaffilmark{6}Kavli Institute for Astrophysics and Space Research, Massachusetts Institute of Technology, 70 Vassar Street, Cambridge, MA 02139, USA}

\begin{abstract}

The quasi-persistent neutron star low-mass X-ray binary \mxb\ went into quiescence in 2001, and we have followed its quiescent X-ray evolution since.  Observations over the first 4 years showed a rapid drop in flux and temperature of the neutron star atmosphere, interpreted as cooling of the neutron star crust which had been heated during the 2.5 year outburst.  However, observations taken approximately 1400 and 2400 days into quiescence were consistent with each other, suggesting the crust had reached thermal equilibrium with the core.  Here we present a new \cxo\ observation of \mxb\ taken 11 years into quiescence and 4 years since the last \cxo\ observation. This new observation shows an unexpected factor of $\sim$3 drop in count rate and change in spectral shape since the last observation, which cannot be explained simply by continued cooling.  Two possible scenarios are that either the neutron star temperature has remained unchanged and there has been an increase in the column density, or, alternatively the neutron star temperature has dropped precipitously and the spectrum is now dominated by a power-law component.  The first scenario may be possible given that \mxb\ is a near edge-on system, and an increase in column density could be due to build-up of material in, and a thickening of, a truncated accretion disk during quiescence. But, a large change in disk height may not be plausible if standard accretion disk theory holds during quiescence.  Alternatively, the disk may be precessing, leading to a higher column density during this latest observation.

\end{abstract}


\section{Introduction}

Neutron stars in transient low-mass X-ray binaries are expected to be hot thermal emitters during quiescence due to pycnonuclear reactions \citep{haensel90} occurring in the deep crust caused by compression during outburst \citep{bbr98}.  The crust, heated during outburst, is then expected to thermally relax once the outburst ends, cooling back into thermal equilibrium with the core \citep{ushomirskyrutledge01, rutledge_ks1731_02}.  This should be particularly noticeable in quasi-persistent transients whose outbursts last years, rather than the more typical weeks to months.  In 2001, two such quasi-persistent transients (\ks\ and \mxb) went into quiescence \citep{wijnands01,wijnands03}, with both showing a significant drop in X-ray flux over the first few years after the end of their outbursts \citep{wijnandsetal02,wijnandsetal04, cackett06, cackett08, cackett10}.  The observed decrease in X-ray flux is consistent with cooling of an accretion-heated neutron star crust.  The rate of cooling and the final  temperature when cooling stops allows us to put constraints on the structure of the crust and state of the core \citep{shternin07, brown09}.  Such crustal cooling has now been observed in 4 additional sources \citep[][Homan et al., in preparation]{degenaar09,degenaarwijnands11,degenaar11a,degenaar11b,fridriksson10,fridriksson11,diaztrigo11}, showing a variety of cooling timescales and temperatures.

Here, we present a new \cxo\ observation of \mxb\ taken in July 2012, approximately 4 years after the last \cxo\ observation and almost 11 years into quiescence.  The first quiescent observation of \mxb\ took place about 1 month into quiescence, finding a thermally dominated spectrum significantly brighter than a ROSAT upper limit from the 1990s \citep{wijnands03}.  Follow-up monitoring showed that \mxb\ cooled rapidly, displaying a factor of 7--9 decrease in X-ray flux in the first 1.5 years \citep{wijnandsetal04}, and a factor of approximately 25 decrease in the first 4 years \citep{cackett06}.  A further \cxo\ observation 6.6 years into quiescence showed that the flux decrease had stopped, indicating the crust was likely back in thermal equilibrium with the core \citep{cackett08}.  In this letter, we discuss a new observation of \mxb\ where we find a significant drop in flux and a change in its spectrum since the previous observation that cannot simply be explained by continued crustal cooling. 

\section{Data Reduction \& Analysis}\label{sec:reduce}
We observed \mxb\ with \cxo\ for approximately 96 ksec at the beginning of July 2012.  This observation was split into two segments.  The first (ObsID: 13711) began on 2012 July 5, lasting 62.2 ksec while the second (ObsID: 14453) began on 2012 July 8, lasting 33.6 ksec.  The ACIS-S instrument was operated using the FAINT data mode, with the source at the nominal aim point.  The data were reduced using the most recent \cxo\ software (\textsc{CIAO} ver. 4.5).  We used the \verb|chandra_repro| tool to reprocess the data with the most recent calibration at the time of the analysis (CALDB ver. 4.5.5.1).  We checked for background flaring and found none.  A circular source extraction region of radius 3\arcsec\ and an annular background extraction region with inner radius of 7\arcsec\ and outer radius of 22\arcsec\ were used.  The source and background spectra and associated response files were extracted using the \verb|specextract| tool.

\mxb\ is a binary system viewed close to edge-on and thus regular X-ray eclipses have been seen from this source during outburst \citep{cominsky84,cominsky89} and quiescence \citep{wijnands03}.  The eclipses last for 900s and occur at the orbital period of 7.1 hours \citep{oosterbroek01}.  Here, the source count rate is too low for us to be able to see eclipses, as well as to clearly distinguish between individual source and background photons.  Therefore, as we have done with previous \cxo\ observations, we used the ephemeris of \citet{oosterbroek01} to determine the times of eclipses during our observation and manually reduce the effective exposure time to compensate for this.  Two eclipses should have occurred during the first segment and one during the second, therefore we reduced the source exposure times by 1800s and 900s giving corrected exposure times of 60.4 ksec and 32.7 ksec, respectively (the background exposure times remain the unchanged).  The average background-subtracted 0.5 -- 10 keV count rate over the two exposures is $(3.2\pm0.7)\times10^{-4}$ counts s$^{-1}$ (note that uncertainties quoted here, and throughout the paper are at the 1$\sigma$ level).  The first exposure has a background-subtracted 0.5 -- 10 keV count rate of $(3.8\pm0.9)\times10^{-4}$ counts s$^{-1}$, while the second exposure has $(2.1\pm1.0)\times10^{-4}$ counts s$^{-1}$, thus both segments have a marginally consistent count rate.

\begin{figure}
\centering
\includegraphics[width=8.5cm]{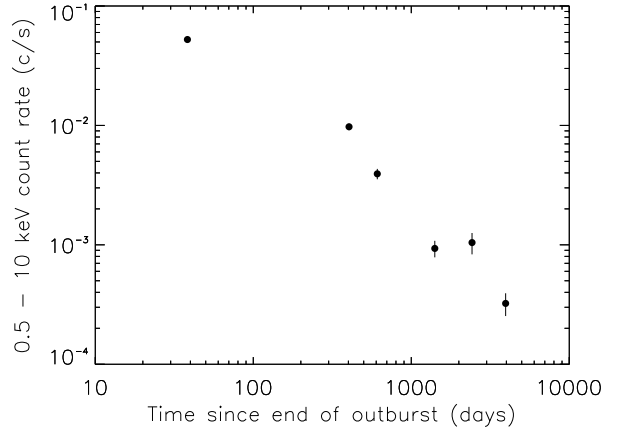}
\caption{The count rate lightcurve of MXB~1659$-$29 as seen during the {\it Chandra} quiescent observations.  Error bars are plotted for all points, but for the first points they are smaller than the symbols.}
\label{fig:rates}
\end{figure}

In Figure~\ref{fig:rates} we compare the count rate of this new observation with the previous 6 \cxo\ observations \citep[see][and references therein]{cackett08}.  Note that we have averaged the count rate from ObsID 5469 and 6337 as the observations were performed only 17 days apart over 1000 days after the end of the outburst.  The count rate of the 2012 observation shows a significant drop from the previous (2008) observation, dropping by a factor of 3 from $(1.0\pm0.2)\times10^{-3}$ counts s$^{-1}$, an approximately 3$\sigma$ difference.  Also note that both segments independently show a significant count rate drop compared to the last observation suggesting that the drop is not a statistical fluctuation.  From Figure~\ref{fig:rates} it is interesting to speculate whether the penultimate observation was anomalously high or whether the latest observation shows an unexpected drop.  Next we detail spectral analysis to try and answer this question.

\subsection{Spectral analysis}

\begin{figure}
\centering
\includegraphics[width=8.2cm]{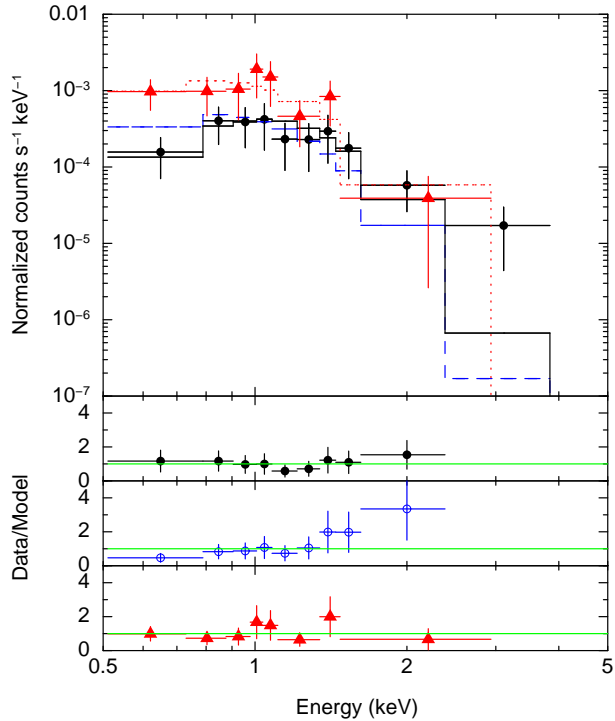}
\caption{{\it Top panel:} {\it Chandra} spectra of MXB~1659$-$29.  Black circles are from the most recent 2012 observation (ObsID: 13711 \& 14453) where the two segments have been added together and rebinned for the purposes of plotting.  The red triangles are from the preceding 2008 {\it Chandra} observation (ObsID: 8984) taken approximately 1500 days earlier.  A significant difference below about 1.5 keV can be seen.  The black solid line is the best-fitting absorbed neutron star atmosphere model when $N_{\rm H}$ is a free parameter, while the blue dashed line is when $N_{\rm H}$ is fixed to the previous value.  The red dotted line is the best-fitting absorbed neutron star atmosphere model to ObsID 8984. There is no significant detection above 4 keV for either observation. {\it Bottom three panels:} These panels show the ratio of the data to the model.  The top ratio panel (black filled circles) shows the ratio of the 2012 data to the model with $N_{\rm H}$ free (the solid black line in the spectrum panel), while the middle ratio panel (blue open circles) shows the ratio of the 2012 data to the model with $N_{\rm H}$ fixed (the dashed blue line in the spectrum panel).  Note that the highest energy bin (approx 2.5 -- 3.5 keV) in these two panels has a ratio of greater than 10 and therefore cannot be seen in this figure.  The bottom ratio panel (red triangles) shows the ratio of the 2008 data to the best-fitting absorbed neutron star atmosphere model (red dotted line in the spectrum panel.)}
\label{fig:spec_compare}
\end{figure}

In order to investigate the change between the observation in 2008 and 2012 we compare the spectra from these two observations.  As can be seen in Figure~\ref{fig:spec_compare}, the most recent observation (black) is significantly fainter compared to the 2008 observation (red) below 1--1.5 keV.  To determine whether this can be explained simply by cooling of the neutron star surface we fit the 2012 observation with an absorbed neutron star atmosphere.  In fitting these data, we fit the spectra from the two segments (ObsID 13711 and 14453) simultaneously, with all model parameters the same for the two segments.  All fitting is performed with \textsc{Xspec} version 12 \citep{arnaud96}. The specific model we fit is phabs$\times$nsa.  We use the neutron star atmosphere model nsa for ease of comparison with our previous analysis in \citet{cackett06} and \citet{cackett08}.  We fix the neutron star radius at 10 km, neutron star mass at 1.4 M$_\odot$ and assume a distance of 10 kpc, which gives an nsa normalization of $1\times10^{-8}$ pc$^{-2}$ \citep[see][for a discussion on the choice of distance and its effects]{cackett08}.  Furthermore, we also assume that the equivalent hydrogen column density, $N_{\rm H}$, remains the same between the observations, fixing it to the  value from \citet{cackett08} of $2.0\times10^{21}$ cm$^{-2}$.  Given the low number of counts we cannot bin the spectra to the minimum number of counts required for use of $\chi^2$ statistics.  We therefore use the C statistic within \textsc{Xspec} to fit the spectra which are binned to a minimum of 1 count per bin.  Note that in the figures we show spectra that have been rebinned for plotting purpose, using the {\sc Xspec} command `setplot rebin 1.6 100'.  While this is a somewhat arbitrary choice, it allows for enough spectral bins to show the shape of the spectrum.

Fitting this simple absorbed neutron star atmosphere model in the 0.5 - 7 keV range we get a best-fitting effective temperature at infinity of $kT_{\rm eff}^{\infty} = 49\pm2$ eV.  However, the model over predicts the count rate below 0.8 keV and under-predicts it above about 1.5 keV, as can be seen in the ratio of the data to the model in Figure~\ref{fig:spec_compare} (blue circles). 
For comparison, the 2008 observation has $kT_{\rm eff}^{\infty} = 56\pm2$ eV.  If, however, we allow the $N_{\rm H}$ to be a free parameter an increased value is found as would be expected given the over-prediction of the flux below 1 keV from the previous model.  This model  leads to $N_{\rm H} = (4.7\pm1.3) \times 10^{21}$ cm$^{-2}$ and $kT_{\rm eff}^{\infty} = 55\pm3$ eV.  This temperature is consistent with the 2008 observation.  Note that if we let the $N_{\rm H}$ value be different for each of the two 2012 segments we get consistent values, however, note that the much shorter second segment has very few counts and thus is not very constraining.

When using the C-statistic, there is not a straightforward reliable way determining the goodness of fit or comparing whether one model is an improvement over another.  We therefore adopt the posterior predictive p-value (ppp) test \citep{protassov02,hurkett08} in order to assess whether one model fits significantly better than another.  This is a Monte Carlo based test to see whether the observed improvement in the C statistic between two different models is significant or not.  For instance, we can compare the fit with the neutron star atmosphere model with $N_{\rm H}$ fixed  with the fit with the neutron star atmosphere model with $N_{\rm H}$ free.  Let us call these model 1 and model 2 respectively for the sake of this discussion.  In the ppp test we simulate 1000 sets of two spectra (one spectrum corresponding to ObsID 13711 and one corresponding to ObsID 14453) using the best fitting model 1.  In simulating the spectra, the parameters for the model are randomly drawn using the covariance matrix of the best-fit as well as using the exposures and background spectra from the real observation.  These 1000 sets of simulated spectra are then fit with both model 1 and model 2 and the best-fitting C-statistic determined for each model.  We then calculate the difference in the best-fitting C-statistic between model 1 and model 2 ($\Delta C$), and define the posterior predictive distribution as the distribution of $\Delta C$ values.  The ppp value is then calculated by comparing the fraction of instances where the simulated $\Delta C$ value is greater than the observed value.  For the case of comparing the neutron star atmosphere model with $N_{\rm H}$ fixed  with the fit with the neutron star atmosphere model with $N_{\rm H}$ free we find that only 13 instances out of 1000 simulations showed a $\Delta C$ value larger than observed, therefore indicating that the model with $N_{\rm H}$ free is better than with $N_{\rm H}$ fixed at the 98.7\% confidence level ($1 - 13/1000 = 0.987$).

Trying an alternative model completely, we fit an absorbed power-law, with the $N_{\rm H}$ fixed at $2.0\times10^{21}$ cm$^{-2}$.  This gives a power-law index of $\Gamma = 2.9\pm0.5$.  This power-law index is softer than would be seen usually in quiescent systems where accretion is on-going \citep[e.g., Cen~X-4,][]{cackett10}.  A ppp test comparing the power-law fit with the neutron star atmosphere fit with $N_{\rm H}$ fixed at the old value shows that the power-law is better at the 99.9\% level (more than 3$\sigma$) while comparing the power-law with the neutron star atmosphere fit with $N_{\rm H}$ free shows that the power-law is better at the 94.3\% confidence level (a little less than 2$\sigma$). Finally, we also test what happens if we try and fit an absorbed neutron star atmosphere plus power-law model.  This is motivated by the fact that the neutron star atmosphere model alone under predicts the count rate above 2 keV (see Fig.~\ref{fig:spec_compare}) possibly indicating the presence of a power-law component in addition to a neutron star atmosphere component.  However, the power-law dominates and the temperature of the neutron star atmosphere goes as low as possible, indicating that two model components is beyond the quality of the data. Alternatively, if we fix the power-law index at $\Gamma = 1.5$ and the $N_H$ at the previous value ($2.0\times10^{21}$ cm$^{-2}$), then we find that the power-law contributes 58\% to the unabsorbed 0.5 -- 10 keV flux and we get $kT_{\rm eff}^{\infty} = 45\pm3$ eV.  If instead we fix $\Gamma = 2$ we then get a power-law fraction of 62\% and $kT_{\rm eff}^{\infty} = 43\pm5$ eV.  Thus, we cannot rule out a significant contribution from a power-law component.  A ppp test comparing the neutron star atmosphere with $N_{\rm H}$ free with the neutron star atmosphere plus power-law model (with $\Gamma = 1.5$) indicates an improvement from adding the power-law at the 94.7\% confidence level. Note that if we fit all the previous observations with an absorbed neutron star atmosphere plus power-law model we find that the power-law component is not statistically required in any of them.  If we also compare the simple power-law fit with the neutron star atmosphere plus power-law fit (with $\Gamma = 1.5$), we find they give equivalent fits with the ppp test indicating no significant improvement (49.9\% confidence level).

\begin{figure}
\centering
\includegraphics[angle=270,width=8.2cm]{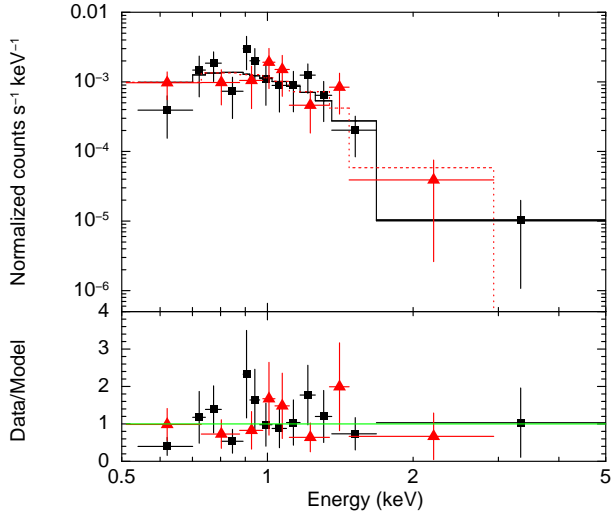}
\caption{{\it Top panel:} {\it Chandra} spectra of MXB~1659$-$29 with best-fitting absorbed neutron star atmosphere model.  Black squares are from the combined 2005 observations (ObsID: 5964 \& 6337) while the red triangles are from the 2008 observation (ObsID: 8984).  The spectra are clearly consistent with one another.  {\it Bottom panel:} The ratio of the data to the best-fitting model shown in the upper panel.}
\label{fig:comp_2005_2008}
\end{figure}

We also compare the 2008 observation with the two observations closely spaced in time taken in 2005, in order to test whether the 2008 observation is anomalously high rather than the 2012 observation being anomalously low.  We show a comparison of the two sets of observations in Figure~\ref{fig:comp_2005_2008}, where we have added the spectra together from ObsID 5469 and 6337 given that they were taken only 17 days apart and have similar spectral shapes.  This figure shows that both observations from 2005 and 2008 look very similar, and hence, it appears to be the 2012 observation which is anomalously low.  The 2008 observation does not appear to show clear signs of a strong power-law component which might be expected from signs of on-going accretion as would be indicated by significant emission above 3 keV.  An increased temperature  caused by on-going accretion \citep[see, for instance, temperature increases in XTE~J1701$-$462 during sporadic accretion events;][]{fridriksson11}  is also not observed.  But, the 2012 observation does appear to be harder than a simple neutron star atmosphere, and is better fit by a simple power-law (a power-law gives an improvement over the best neutron star atmosphere model at the 94.3\% confidence level).  This may possibly indicate some cooling between 2012 and 2008, though the extremely low count rate limits what constraints we can get on the temperature from the 2012 observation when fitting with a neutron star atmosphere plus power-law model.

Finally, we rule-out that there has been a build-up of contaminant on the optical blocking filters \citep[see, e.g.][]{marshall04} between ObsIDs 8984 and 13711/14453 that is yet to be modeled and included in the detector responses.  We extract the spectrum from nearby source CXOU~J170205.9$-$295619 from ObsIDs 8984 and 13711/14453  using \verb!specextract!.  We find that this source has a constant count rate and spectral shape (absorbed power-law) between the observations. In particular it shows no signs of any changes below 1.5 keV.

\section{Discussion}\label{sec:discuss}

We have continued to monitor the spectrum of the crustal cooling source \mxb\ since it entered into quiescence in 2001.  Here, we have presented a new \cxo\ observation from July 2012 (almost 11 years into quiescence).  While initial monitoring of \mxb\ showed that it cooled rapidly over the first few years in quiescence, the observation from 2008 indicated that cooling had appeared to halt \citep{cackett08}.  However, this new observation shows an unexpected drop in count rate by a factor of 3 since the preceding \cxo\ observation performed 4 years earlier.   Spectral analysis comparing the new 2012 data with the 2008 spectrum shows that the 2012 spectrum is systematically below the 2008 spectrum at energies less than about 1--1.5 keV (Figure~\ref{fig:spec_compare}).

We are left with several options to explain the count rate drop between the 2008 and 2012 observations.  Firstly, the drop in count rate is not real, but a statistical outlier.  However, note that the drop is at the 3$\sigma$ level and that both separate pointings in 2012 show a similar count rate drop.  Secondly, the neutron star has cooled slightly, but the $N_{\rm H}$ has remained constant.  Thirdly, the neutron star has remained at a constant temperature but the $N_{\rm H}$ has increased.  Fourthly, the neutron star temperature has dropped precipitously and the spectrum is now dominated by a (rather soft) power-law.  Option two is not favored based on spectral fitting - a model with temperature free and $N_{\rm H}$ remaining constant shows clear residuals (see Fig.~\ref{fig:spec_compare}).  Options three and four both give improved spectral fits.   For instance, the fit is improved (at the 98.7\% confidence level) by fitting an absorbed neutron star atmosphere where the $N_{\rm H}$ has increased from $2.0\times10^{21}$ cm$^{-2}$ \citep{cackett08} to $N_{\rm H} = (4.7\pm1.3) \times 10^{21}$ cm$^{-2}$ yet the neutron star temperature is consistent with remaining constant.  Alternatively, the spectrum is well fit by an absorbed power-law with index $\Gamma = 2.9\pm0.5$ with an unchanged $N_{\rm_H}$.

Let us first examine the possibility of an increase in column density.  In this scenario, the neutron star is back in thermal equilibrium as we see that the temperature is consistent between the 2008 and 2012 observations.  But, we need a physical motivation to explain the increase in column density -- in most systems a change in column density would not be expected.  However, \mxb\ is a nearly edge-on system that shows eclipses both in outburst and quiescence.  Models for the accretion disk during quiescence usually have the disk truncated at a few thousand Schwarzschild radii, essentially acting as a reservoir for material transferred from the companion star \citep[e.g.][]{lasota96,esin97,dubus01}.  Material then builds up in the outer disk during quiescence until an outburst is triggered.  At the time of the 2012 observation, \mxb\ was almost 11 years into quiescence.  The previous known quiescent period for this object is 21 years \citep{intzand99}.  If \mxb\ follows a similar pattern this time, then it is approximately half way through its quiescent period.  We can speculate, then, that during quiescence the build-up of material could increase the scale height of the outer disk, leading to an increase in absorption.  Under such a scenario, we might expect the absorbing column to further increase as quiescence progresses, which can be tested by future observations.

Considering standard accretion disk theory, however, a large change in scale height during quiescence is not expected.  If we assume that the standard $\alpha$ disk holds during quiescence, then we can determine how the disk properties scale.  In order that the truncated disk remains stable, and an outburst is not triggered, the temperature in the outer disk cannot rise by a large factor.  Yet, for a given radius, the height of the disk scales like $c_s \sim \sqrt{T}$ (where $c_s$ is the sound speed and $T$ the temperature in the disk).  Thus, a factor of 2 increase in temperature would only lead to a 40\% increase in the disk height.  Moreover, the temperature in quiescence is also likely to be colder than when in outburst, thus, a lower scale height would be expected in quiescence.  Hence, if standard disk theory holds in quiescence a significant change in disk height may not be plausible.

An alternative explanation for the increase an increase in column density is that the accretion disk could be precessing.  In this scenario, during the 2012 observations our line of sight passes through more of the disk than in previous observations.  Superorbital periods have been detected in at least 25 X-ray binaries \citep[see][and references therein]{kotzecharles12}.  The two main mechanisms for superorbital periods are precession \citep{whitehurst91} or radiation-induced warping \citep{ogilvie01}, though see \citet{kotzecharles12} for suggestions of other possible mechanisms.  During quiescence, irradiation of the outer disk will be many orders of magnitude less than during outburst, thus radiation-induced warping would not be expected.  Precession, however, may occur when there is an extreme mass ratio when $M_2/M_1 \leq 0.25-0.33$ \citep{whitehurst91}. \mxb\ has a known optical counterpart \citep{wachter00, wijnands03}, which is thought to be a K0 star \citep{wachter98, wachter00}.  Such a companion, however, would not lead to the required mass ratio for precession.  No significant detection of a superorbital period has been reported for \mxb, though \citet{kotze12} report a very marginal detection at the 1.2$\sigma$ level at a period of about 350 days.  Thus, we cannot completely rule out the possibility of precession, despite the mass ratio.  In this precession scenario, we would more likely than not see a return to the previous column density in future observations, though this depends on the exact modulation timescale for the precession. We may also expect to have already seen some hints of this in previous data, but again, this depends on the precession period and the sparse sampling of X-ray observations we have may not be enough to have detected it.

\begin{figure}
\centering
\includegraphics[width=8.5cm]{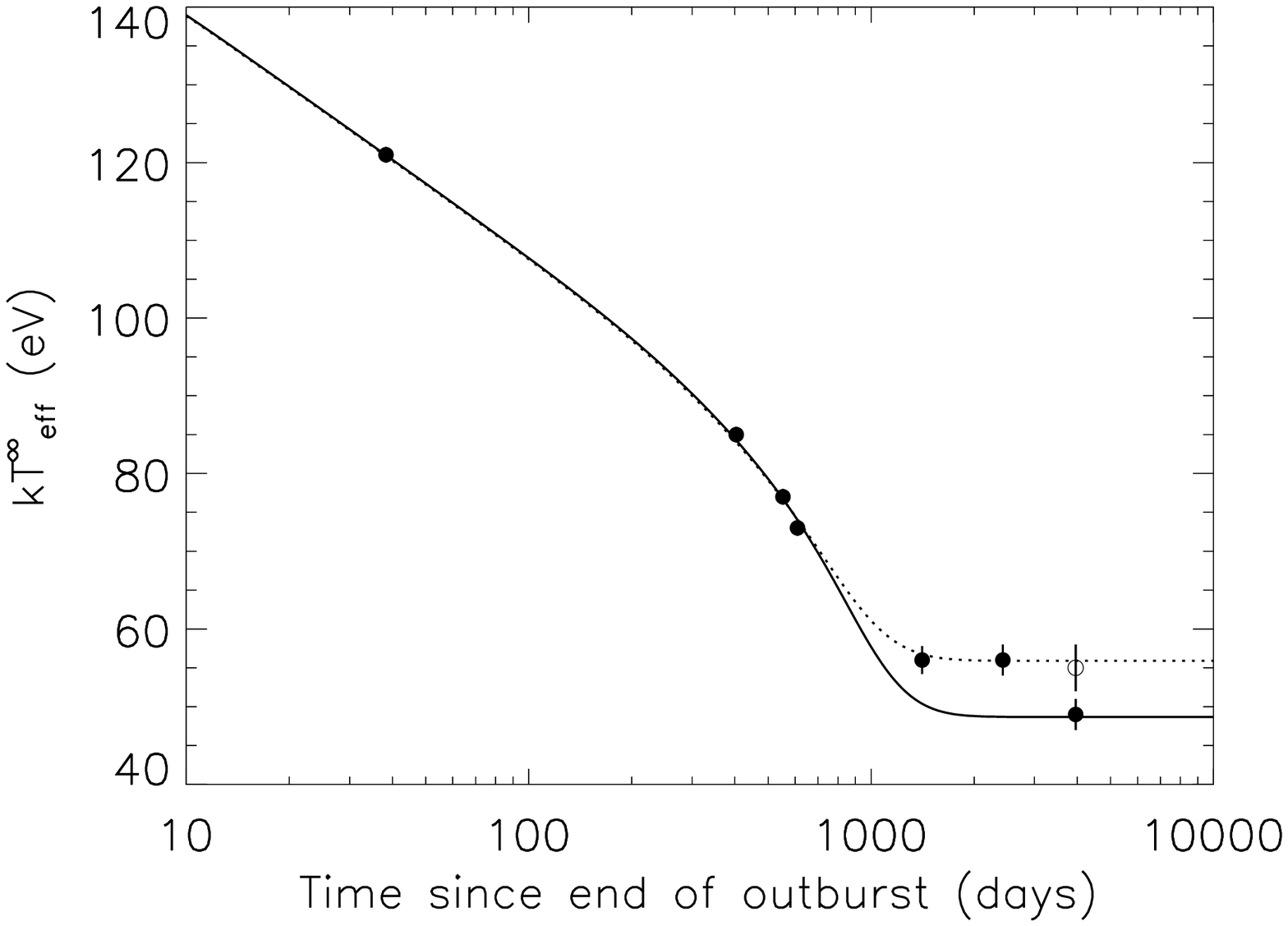}
\caption{Theoretical cooling curves calculated using the model described in \citet{brown09} compared to the observed temperatures \citep[though note ObsIDs 5469 and 6337 have been averaged]{cackett08}.  The dotted line shows the best-fitting model from \citet{brown09}, while the solid line shows the best-fitting model if a temperature of 49 eV is assumed for the 2012 observation (the fit with $N_{\rm H}$ fixed, filled circle).  The additional point causes a poor fit at times between 1000 and 3000 days.  Note that the final data point becomes consistent with the dotted line if the fit with $N_{\rm H}$ as a free parameter is used (open circle).}
\label{fig:model}
\end{figure}

We now consider the other possibility, that the spectrum is now dominated by a power-law component.  This would require that the neutron star surface cooled dramatically between 2008 and 2012.  Unfortunately, the data quality does not permit constraints on the neutron star temperature, as a second component (in addition to the power-law) is not required by the data.  But, if dominated by the power-law then such a component could be due to low-level accretion during quiescence or alternative mechanisms.   However, there is no sign of a strong power-law component in previous observations (see Fig.~\ref{fig:comp_2005_2008}), and, the 2008 and 2012 spectra are very similar above about 1.5 keV.  Previous observations can be fit by a single absorbed power-law, but such fits give a high power-law index, e.g. ObsID 8984 gives $\Gamma = 4.2\pm0.5$. Additionally, in the presence of low-level accretion one would expect a somewhat warmer neutron star component \citep[e.g.][]{fridriksson11}, while here we see a cooler one.  Moreover,  the photon index of approximately 3 is softer than other quiescent neutron stars, especially those that just display a power-law and no clear thermal component, such as EXO~1745$-$248 \citep{wijnands05} and SAX J1808.4$-$3658 \citep{heinke09}.  Signatures of on-going accretion might include sporadic variability in quiescence as has been seen in several other quiescent neutron stars \citep{campana97,campanaetal04, rutledgeetal01a, rutledge02, cackett05, cackett10,cackett11,fridriksson11}, and could be seen via future observations. 

If we consider that the spectrum is now power-law dominated, then we must discuss the implications for crustal cooling.  A significant drop in temperature cannot be reconciled with the crust cooling models of \citet{brown09}.  Even a modest drop to just 49 eV (as implied by the spectral fit with an absorbed neutron star atmosphere and fixed column density) causes the model to fit the 2005 and 2008 data points poorly, as we show in Figure~\ref{fig:model}.  \citet{pagereddy12} do show a model (their figure 12) which shows a further decline in temperature at late times.  However, this model is calculated specifically for XTE~J1701$-$462, which was accreting above the Eddington limit during an outburst of only 1.6 years \citep{homan07,lin09}.  This high accretion rate for a comparatively short time leads to a crust temperature profile with two peaks - one near the surface, and one in the inner crust, with a dip in between. This leads to a cooling curve with an initial drop that plateaus (corresponding to the dip in the crust profile) and then has a second drop at late times.    For a longer outburst at a lower accretion rate (as is the case for \mxb) the temperature profile of the crust should be flatter (without two separate peaks), and thus the a second drop in temperature is not expected for \mxb.

In conclusion, we cannot definitively explain the cause of the drop in count rate between the 2008 and 2012 observation of \mxb.  However, the possible scenarios predict different future behavior.  The picture where material is building up in the outer disk and leading to an increase in absorption leads to the expectation that a further increase in column density could be seen in the future.  In the precession scenario we would most likely expect the $N_{\rm H}$ to return to its previous value in the future. Alternatively, the explanation involving low-level accretion and a much colder neutron star leads to the expectation of sporadic variability in quiescence.  Future observations of \mxb\ can test between these scenarios.

\acknowledgements
We thank Caroline D'Angelo for helpful discussions, and we thank the referee for helpful suggestions that have improved the paper.  Support for this work was provided by NASA through Chandra Award Number GO13047X issued by the Chandra X-ray Observatory Center.   N.D. is supported by NASA through Hubble postdoctoral fellowship grant number HSTHF-51287.01-A from the Space Telescope Science Institute. RW is partially supported by a European Research Council Starting Grant.

\end{document}